\documentclass[conference]{IEEEtran}
\usepackage{amsmath,amssymb,bm,amsthm}
\usepackage{graphicx,subfig}

\ifCLASSINFOpdf
\else
\fi
\begin{document}
%
\title{Optimal Power Allocation over Multiple Identical Gilbert-Elliott Channels}


\author{\IEEEauthorblockN{Jiaming Li}
\IEEEauthorblockA{School of Information Security\\
Engineering\\
Shanghai Jiao Tong University, China\\
Email: jljl@sjtu.edu.cn}
\and
\IEEEauthorblockN{Junhua Tang}
\IEEEauthorblockA{School of Information Security\\
Engineering\\
Shanghai Jiao Tong University, China\\
Email: junhuatang@sjtu.edu.cn}
\and
\IEEEauthorblockN{Bhaskar Krishnamachari}
\IEEEauthorblockA{Ming Hsieh Department\\
of Electrical Engineering\\
Viterbi School of Engineering\\
University of Southern California\\
Email: bkrishna@usc.edu}
}


%


\maketitle

\begin{abstract}
We study the fundamental problem of power allocation over multiple Gilbert-Elliott communication channels. In a communication system with time varying channel qualities, it is important to allocate the limited transmission power to channels that will be in good state. However, it is very challenging to do so because channel states are usually unknown when the power allocation decision is made. In this paper, we derive an optimal power allocation policy that can maximize the expected discounted number of bits transmitted over an infinite time span
by allocating the transmission power only to those channels that are believed to be good in the coming time slot. We use the concept \emph{belief} to represent the probability that a channel will be good and derive an optimal power allocation policy that establishes a mapping from the channel \emph{belief} to an allocation decision. Specifically, we first model this problem as a partially observable Markov decision processes (POMDP), and analytically investigate the structure of the optimal policy. Then a simple threshold-based policy is derived for a three-channel communication system. By formulating and solving a linear programming formulation of this power allocation problem, we further verified the derived structure of the optimal policy.

\end{abstract}


%
\IEEEpeerreviewmaketitle


\section{Introduction}

Communication over the wireless medium is subject to multiple impairments such as fading, path loss, and interference. These effects degrade the quality of received signal and lead to transmission failures. The quality of the radio channel is often random and evolves in time, ranging from good to bad depending on the propagation conditions. To cope with the changing channel quality and achieve a better channel utilization, it is important to adopt link adaptation schemes whereby data/coding rate and transmit power of the transmitted signal are adaptively adjusted according to the channels conditions \cite{cite1,cite2,cite3,cite4}.

Adaptive power control is an important technique to select the transmission power of a wireless system according to channel condition to achieve better network performance in terms of higher data rate or spectrum efficiency [1],[2]. There has been some recent work on power allocation over stochastic channels \cite{Zaidi,Xwang,YGai}, but the problem of optimal power allocation
across multiple dynamic stochastic channels is challenging and remains largely unsolved from a theoretical perspective.

We consider a wireless communication system operating on $N (N \geq 3$) parallel transmission channels. Each channel is modeled as a time slotted two-state Markov model known as the Gilbert-Elliot channel. This model assumes that the channel can be in either a good state or a bad state. The channel in a good state can transmit at a certain rate successfully but a channel in bad state will lead to transmission failure and therefore suffer data loss. We assume all channels in the system are statistically identical and independent of each other.  Our goal is to allocate the total transmission power only to channels in good state so as to maximize the expected discounted number of bits transmitted over an infinite time span. Since the channels sates are unknown at the time this power allocation decision is made, this problem is more challenging than it looks like.

There have already been some related works on the decision-making problem over Gilbert-Elliott channels in the literature. In \cite{cite3} and \cite{cite4}, the authors used Markov Decision Process (MDP) tools to establish an optimal threshold strategies that minimize the transmission consumption and maximize the throughput over one Gilbert-Elliott channels. In \cite{cite5}, the authors defined three transmitting actions and solved the problem of dynamically choosing one of them to maximize the expected discounted number of bits transmitted. In \cite{cite6}, the authors study the problem of choosing a transmitting strategy from two choices emphasizing the case when the channel transition probabilities are unknown. The work in \cite{cite7} and \cite{cite8} is most relevant to the work in this paper, the differences between these three are as follows: \cite{cite7} addresses power allocation problem in the context of two identical channels and three allocation strategies: betting on channel 1, betting on channel 2 and using both channels, whilst \cite{cite8} added one more action of using none of the channels and introduced penalty caused by transmission on a bad channel. The spirit of this paper is similar to those in \cite{cite7} and \cite{cite8}, but addresses a more challenging setting involving $N$ identical channels($N \geq 3$). When $N$ is large, the power allocation decisions becomes much more complicated, and it is more difficult to derive and express the optimal policy.

In this paper, we formulate our power allocation problem as a partially observable Markov decision process (POMDP). We then treat the POMDP as a continuous state MDP and develop the structure of the optimal policy (decision). Our main contributions are summarized as follows: (1) we formulate the problem of dynamic power allocation over multiple parallel Gillber-Elliott channels using the MDP theory, 2) we theoretically prove some key properties of the optimal policy for this particular problem, and derive the exact optimal policy for the three-channel system, (3) through simulation based on linear programming, we verify the structure of the optimal policy and demonstrate how to numerically compute the thresholds and construct of the optimal policy when system parameters are known.

\section{Problem Formulation}
\subsection{Channel model and assumptions}

In this paper, we consider a wireless communication system operating on $N$ parallel channels. We assume that these channels are statistically identical and independent of each other. Each channel is modeled by a time slotted Gilbert-Elliott channel which is a one dimensional two-state Markov chain $G_{i,t}(i \in \{1,2,...,N\},t \in \{1,2,...,\infty\})$ (\begin{math}i\end{math} is the index of channel and $t$ is time slot). $G_{i,t}=1$ means the channel is in good state in time slot $t$, and $G_{i,t}=0$ means the channel is in bad state in time slot $t$. The state transition probability is denoted by: $Pr[G_{i,t}=1|G_{i,t-1}=1]=\lambda_1$ and $Pr[G_{i,t}=1|G_{i,t-1}=0]=\lambda_0,i \in \{1,2,...,N\}.$ We assume the state transitions happen at the beginning of each time slot and share a positive correlation assumption that $\lambda_0 \leq \lambda_1$ which means the probability of retaining in good state is higher than that of recovering from a bad state.

The total transmitting power of the communication system is $P$. At the beginning of each time slot, system needs to allocate the limited power to the channels optimally. Let $P_i(t)$ denote the power allocated to channel $i$ at time $t$, we have:
\begin{equation}
P=\sum_{i=1}^N{P_i(t)} \label{eq:eq1}.
\end{equation}

We assume that the states of channels are unknown at the beginning of each time slot. If channel $i$ is used in time slot $t$ ($P_i (t)>0$), the state of channel $i$ in slot $t$ is revealed at the end of that slot through a feedback mechanism. Otherwise, if channel $i$ is not used ($P_i (t)=0$), its exact state during time slot $t$ remains unknown. Therefore this power allocation problem is challenging because decisions have to be made when current channel states are unknown.

 To simplify the problem, we adopt the following power allocation strategies. At the beginning of each time slot, the system chooses $k$ (hopefully good channels) out of the $N$ channels and allocates total power $P$ to the $k$ channels equally. So each of the selected channel is allocated $P/k$ of the transmission power. If a channel is allocated $P/k$ power, there are two different consequences: 1) the channel is in good state and sends $R_k (k \leq N)$ bits of data successfully (reward); 2) the channel is in bad state and suffers $C_k (k \leq N)$ bits of data loss due to poor channel quality (penalty). We assume that $R_{k_2}<R_{k_1}<\frac{k_2}{k_1}R_{k_2}$, $C_{k_2}<C_{k_1}<\frac{k_2}{k_1}C_{k_2}$ $(1 \leq k_1 \leq k_2 \leq N)$. For all $1 \leq k \leq N$, we have $R_k>C_k$. If a channel is not allocated any transmission power, it has zero reward and zero penalty.

 We define an n-dimensional vector $\bm{\alpha}_i\!=\!(a_{i,1},a_{i,2},...,a_{i,N})$ to denote allocation action $i$, where $1 \leq i \leq 2^N$, $a_{i,j} \in \{0,1\}$, where $a_{i,j}=1$ means channel $j$ is used in action $i$ and $a_{i,j}=0$ means channel $j$ is not used in this action. Because the total number of channels is $N$ and each channel can be either used or not, there are $2^N$ possible allocation actions. We use $\mathbf{B}=\{\bm{\alpha}_i, i \in {1,2,...,2^N}\}$ to denote the set of all $2^N$ different allocation actions. Define ${||\bm{\alpha}_i||}=a_{i,1}^2+a_{i,2}^2+...+a_{i,N}^2=a_{i,1}+a_{i,2}+...+a_{i,N}=k$ as the number of used channels in this action. ((When $k$ is large, the system spreads the risk of data loss to more channels and is more likely to get a mediocre reward. When $k$ is small, the system bets on less channels and might lead to better reward. The focus of this paper is to find an optimal allocation policy that maximizes the long term discounted reward.))

\subsection{Formulation of the Partially Observable Markov Decision problem}

\indent As described above, at the beginning of each time slot, the system needs to choose an appropriate strategy $\alpha_{i}$  in order to maximize the data transmitted in the long term. Due to the fact that the exact channel state is not observable when this decision is made, this problem can be described as a Partially Observable Markov Decision Process (POMDP). In \cite{cite9}, it is shown that given the past history, a sufficient statistic for determining the optimal policy is the conditional probability that the channel is in the good state at the beginning of the current time slot which is called the belief. We denote the belief by a N-dimensional vector $\mathbf{x}_t=(x_{1,t},x_{2,t},...,x_{N,t})$, where $x_{i,t}=Pr[G_{i,t}=1|\hbar_t ]$, $i \in \{1,2,...,N\}$, $\hbar_t$ is all the history before time slot $t$. By introducing the belief, we can convert the POMDP into a Markov Decision Process (MDP) with an uncountable state space $\mathbf{O}=(\underbrace{[0,1],[0,1],...,[0,1]}_N)$.

\indent Define policy $\pi$ as the decision-making rules which is a mapping from the state space $\mathbf{O}$ to the actions space $\mathbf{B}$. Define $V^\pi (\mathbf{p})$ as the expected discounted number of data transmitted with initial belief $\mathbf{p}=(p_1,p_2,...,p_N)$, where $p_i=Pr[G_{i,0}=1|\hbar_0]=x_{i,0},\ i \in \{1,2,...,N\}.$ We have:
\begin{equation}
V^\pi(\mathbf{p})=E^\pi[\sum_{t=0}^{\infty}{\beta^tg_{a_t}(\mathbf{x}_t)|\mathbf{x}_0=\mathbf{p}}]  \label{eq:eq2}
\end{equation}
where $E^\pi$ is the expectation given policy $\pi$, $\beta$ is the discount factor, $t$ is time slot, $a_t\in \mathbf{B}$  denotes the action taken in time slot $t$, and $g_{a_t}(\mathbf{x}_t)$ denotes the expected immediate reward when choosing action $a_t$ given the belief $\mathbf{x}_t$. Let $\bm{\alpha}_{i_t}$ denote the action $a_t$, then $||\bm{\alpha}_{i_t}||$ is the number of channels used in this action, we have:
\begin{equation}
g_{a_t}(\mathbf{x}_t)=\sum_{j=1}^N{a_{i_t,j} x_{j,t}(R_{||\bm{\alpha}_{i_t}||}+C_{||\bm{\alpha}_{i_t}||})-||\bm{\alpha}_{i_t}||C_{||\bm{\alpha}_{i_t}||}} \label{eq:eq3}
\end{equation}
Let set $\mathbf{m_t}=\{m_1,m_2,...,m_{||\bm{\alpha}_{i_t}||}\}, m_i \in \{1,2,...,N\}$ be the set of channels chosen by action $\bm{\alpha}_{i_t}$, $m_i \neq m_j \ (i \neq j)$, equation (\ref{eq:eq3}) can be rewritten as:
\begin{equation}
g_{a_t}(\mathbf{x}_t)=\sum_{j \in \mathbf{m}_t}{x_{j,t}(R_{||\bm{\alpha}_{i_t}||}+C_{||\bm{\alpha}_{i_t}||})-||\bm{\alpha}_{i_t}||C_{||\bm{\alpha}_{i_t}||}} \label{eq:eq4}
\end{equation}
Now we define the value function $V(\mathbf{p})$ as:
\begin{equation}
V(\mathbf{p})=\max_{\pi}{\,V^\pi(\mathbf{p})} \quad \forall \ \mathbf{p} \in \mathbf{O} \label{eq:eq5}
\end{equation}
A policy is called stationary if it is a function mapping the state space $\mathbf{O}$ to action space $\mathbf{B}$. It is proved that there exists a stationary policy $\pi^*$ that satisfies $V(\mathbf{p})=V^{\pi^*}(\mathbf{p})$ and also the Bellman equation \cite{cite10}:
\begin{equation}
V(\mathbf{p})=\max_{a \in \mathbf{B}}{\,\{V_a(\mathbf{p})\}} \label{eq:eq6}
\end{equation}
where $V_a(\mathbf{p})$ denotes the value acquired when the belief is $\mathbf{p}$ and the immediate action is $a$:
\begin{equation}
V_a(\mathbf{p})=g_a(\mathbf{p})+\beta E^\mathbf{y}[V(\mathbf{y})|\mathbf{x}_0=\mathbf{p}, a_0=a] \label{eq:eq7}
\end{equation}
where $\mathbf{y}$ denotes the belief at the beginning of next time slot when action $a$ is taken, $E^{\mathbf{y}}$ denotes the expectation of total reward when the belief of next time slot is $\mathbf{y}$.

Next we discuss the expression of $V_a(\mathbf{p})$. For each action $a=\bm{\alpha}_i \in \mathbf{B}$, there are two types of channels: used and unused. For a used channel $j$, it is allocated $P/||\bm{\alpha}_{i}||$ transmission power, thus it will have immediate reward $p_jR_{||\bm{\alpha}_{i}||}$ and immediate loss $(1-p_j)C_{||\bm{\alpha}_{i}||}$. Since the channel state in the current time slot is revealed at the end of this time slot through feedback, the belief of channel $j$ in the next time slot will be either $\lambda_1$ (if channel $j$ is in good state in the current time slot) or $\lambda_0$ (if channel $j$ is in bad state in the current slot).

For any unused channel $j$, there will be no immediate reward or loss, and there is no feedback to reveal the channel state. Therefore, the belief in the next time slot is calculated as:
\begin{equation}
T(p_j)=(1-p_j)\lambda_0+p_j\lambda_1=\sigma p_j+\lambda_0 \label{eq:eq8}
\end{equation}
where $\sigma=\lambda_1-\lambda_0$.

For ease of notation, we omit the subscript $i$ of $\bm{\alpha}_i$ and use $\bm{\alpha}$ to denote a certain action taken in a certain time slot in the following discussions. Let $\mathbf{m}=\{m_1,m_2,...,m_{||\bm{\alpha}||}\}, m_k \in \{1,2,...,N\}$ be the set of channels used in action $\bm{\alpha}$. Let $\bm{\varphi}_i=(\varphi_{i,1},\varphi_{i,2},...,\varphi_{i,||\bm{\alpha}||}), \varphi_{i,k} \in \{0,1\}, k \in \{1,2,...,||\bm{\alpha}||\}$ denote the state of the used channels in the elapsed time slot. Since each of the used channel may be in good or bad state, the total number of possible states of the $||\bm{\alpha}||$ used channels is $2^{||\bm{\alpha}||}$, and we use $\bm{\Psi}=\{\bm{\varphi}_i | i=1,...,2^{||\bm{\alpha}||}\}$ to denote the set of all possible states of used channels. For the convenience of notation, we represent the probability of state $\bm{\varphi}_i$ as
\begin{equation}
f(\bm{\varphi}_i)=\prod_{k=1}^{||\bm{\alpha}||}h(\varphi_{i,k})\label{eq:eq9}
\end{equation}
where
\begin{equation}
h(\varphi_{i,k})=\begin{cases}
 p_{m_k} & \text{ if } \varphi_{i,k}=1  \\
 1-p_{m_k} & \text{ if } \varphi_{i,k}=0
\end{cases} \label{eq:eq10}
\end{equation}
For each $\bm{\varphi}_i$, the corresponding system belief in the next time slot is $\mathbf{y}_{\bm{\varphi}_i}^*=(y_1^*,y_2^*,...,y_N^*)$, where
\begin{equation}
y_j^*=\begin{cases}
 \lambda_0 & \text{ if } j = m_k \text{ and } \varphi_{i,k}=0  \\
 \lambda_1 & \text{ if } j = m_k \text{ and } \varphi_{i,k}=1  \\
 T(p_j)    & \text{ otherwise }
\end{cases} \label{eq:eq11}
\end{equation}
From (\ref{eq:eq9})-(\ref{eq:eq11}), we know that the belief of next time slot will be $\mathbf{y}_{\bm{\varphi}_i}^*$ with the probability of $f(\bm{\varphi}_i)$. So the conditional value function $V_{\bm{\alpha}}(\mathbf{P})$ is calculated as:
\begin{eqnarray}
V_{\bm{\alpha}}(\mathbf{p})&=&\sum_{k=1}^{||\bm{\alpha}||}{p_{m_k}(R_{||\bm{\alpha}||}+C_{||\bm{\alpha}||})}-{||\bm{\alpha}||}C_{||\bm{\alpha}||}  \nonumber\\
&&+\beta\sum_{\bm{\varphi}_i \in \bm{\Psi}}{f(\bm{\varphi}_i)V(\mathbf{y}_{\bm{\varphi}_i}^*)}
\label{eq:eq12}
\end{eqnarray}
More specifically, the last term of (\ref{eq:eq12}) can be written as
\begin{eqnarray}
&  &\sum_{\bm{\varphi}_i \in \bm{\Psi}}{\!\!f(\bm{\varphi}_i)V(\mathbf{y}_{\bm{\varphi}_i}^*)}\nonumber\\
&\!\!\!\!\!=&\!\!\!\!\!(1-p_{m_1})\cdots(1-p_{m_M})V(\mathbf{y}_{(0,0,...,0)}^*)\nonumber\\
&\!\!\!\!\!+&\!\!\!\!\!p_{m_1}(1-p_{m_2})\cdots(1-p_{m_M})V(\mathbf{y}_{(1,0,...,0)}^*)\nonumber\\
&\!\!\!\!\!+&\!\!\!\!\!(1-p_{m_1})p_{m_2}\cdots(1-p_{m_M})V(\mathbf{y}_{(0,1,...,0)}^*)\nonumber\\
&\!\!\!\!\!+&\!\!\!\!\!\cdots\nonumber\\
&\!\!\!\!\!+&\!\!\!\!\!(1-p_{m_1})(1-p_{m_2})\cdots p_{m_M}V(\mathbf{y}_{(0,0,...,1)}^*)\nonumber\\
&\!\!\!\!\!+&\!\!\!\!\!p_{m_1}p_{m_2}\cdots(1-p_{m_M})V(\mathbf{y}_{(1,1,...,0)}^*)\nonumber\\
&\!\!\!\!\!+&\!\!\!\!\!\cdots\nonumber\\
&\!\!\!\!\!+&\!\!\!\!\!(1-p_{m_1})\cdots p_{m_{M\!-\!1}}p_{m_M}V(\mathbf{y}_{(0...,1,1)}^*)\nonumber\\
&\!\!\!\!\!+&\!\!\!\!\!\cdots\nonumber\\
&\!\!\!\!\!+&\!\!\!\!\!p_{m_1}p_{m_2}\cdots p_{m_M}V(\mathbf{y}_{(1,1,...1)}^*)
\label{eq:eq13}
\end{eqnarray}
where $M=||\bm{\alpha}||$. The Bellman equation (\ref{eq:eq6}) can then be expressed as:
\begin{equation}
V(\mathbf{p})=\max_{\bm{\alpha}}{V_{\bm{\alpha}}(\bm{p})}\label{eq:eq14}
\end{equation}

\section{Structure of the Optimal Policy}
In this section, we will first study the structural features of the optimal policy, and then derive the optimal policy for power allocation over three identical channels.
\subsection{Properties of value function}
\indent \textbf{\emph{Lemma 1:}} The value function $V_a(\mathbf{p}), a \in \mathbf{B}$ is affine in $p_j$ and the following equality holds:
\begin{eqnarray}
&&V_a(p_1,p_2,...,p_{j-1},cp+(1-c)p',p_{j+1},...,p_N)\nonumber\\
&=&cV_a(p_1,p_2,...,p_{j-1},p,p_{j+1},...,p_N) \nonumber\\
&+&(1-c)V_a(p_1,p_2,...,p_{j-1},p',p_{j+1},...,p_N)
\label{eq:eq15}
\end{eqnarray}
where $0 \leq c \leq 1$ is a constant, $j \in \{1,2,...,N\}$. In this paper we use the following definition of ``affine'': $h(x)$ is said to be affine with respect to $x$ if $h(x)=ax+c$ with constant $a$ and $c$.
\begin{proof}
The equality in (\ref{eq:eq15}) naturally holds if $V_a(\mathbf{p}), a \in \mathbf{B}$ is affine in $p_j$ for all $j$. So we only need to prove the first half of the lemma.

Suppose the system chooses action $a=\bm{\alpha}$ in a certain time slot. Let $M=||\bm{\alpha}||$ be the number of used channels, $\mathbf{m}=\{m_1,m_2,...,m_M\}, m_j \in \{1,2,...,N\} (j=1,...,M)$ be the set of channels chosen by action $\bm{\alpha}$. First we prove that Lemma 1 is true for used channels in $\mathbf{m}$. It is clear from equation (\ref{eq:eq12}) that the first term on the right side of equation (\ref{eq:eq12}) is affine in $p_{m_j}(j=1,...,M)$, and from equation (\ref{eq:eq13}) it is clear that the last term on the right side of equation (\ref{eq:eq12}) is also affine in $p_{m_j}(j=1,...,M)$. Therefore we say that for each used channel $j$( $j \in \mathbf{m}$), the value function $V_a(\mathbf{p})$ is affine in $p_j$.

Next we need to prove that $V_a(\mathbf{p})$ is also affine in $p_j$ for unused channel $j(j \notin \{m_1,m_2,...,m_M\})$. From equation (\ref{eq:eq12}), we can see that the first and second terms on the right side of the equation do not have the term $p_j\:(j \notin \{m_1,m_2,...,m_M\})$, so we just need to consider the third term $\beta\sum_{\bm{\varphi_i} \in \bm{\Psi}}{f(\bm{\varphi_i})V(\mathbf{y}_{\bm{\varphi_i}}^*)}$. From equation (\ref{eq:eq12}) and (\ref{eq:eq13}), we know that if $V(\mathbf{y}_{\bm{\varphi_i}}^*)$ is affine in $p_j$, the lemma holds.

\indent From (\ref{eq:eq14}), we know $V(\mathbf{y}_{\bm{\varphi}_i}^*)=V_{\bm{\alpha}'}(\mathbf{y}_{\bm{\varphi}_i}^*)$, where $\bm{\alpha}'$ is the optimal action to maximize $V(\mathbf{y}_{\bm{\varphi}_i}^*)$. If channel $j$ is used in action $\bm{\alpha}'$, then according to (\ref{eq:eq12}) and the fact that $T(p_j)=(1-p_j)\lambda_0+p_j\lambda_1=\sigma p_j+\lambda_0$ is affine in $p_j$, we can say $V(\mathbf{y}_{\bm{\varphi}_i}^*)=V_{\bm{\alpha}'}(\mathbf{y}_{\bm{\varphi}_i}^*)$ is affine in $p_j$. If channel $j$ is not chosen in action $\bm{\alpha}'$, we have $y_j^*=T(p_j)$, then $V(\mathbf{y}_{\bm{\varphi}_i}^*)$ can be expressed as:
\begin{eqnarray}
V(\mathbf{y}_{\bm{\varphi}_i}^*)&\!\!\!\!=&\!\!\!\!V_{\bm{\alpha}'}(y_1^*,y_2^*,...,T(p_j),...,y_N^*)\nonumber\\
&\!\!\!\!=&\!\!\!\!\sum_{k=1}^M{y_{m_k'}^*(R_M+C_M)}-MC_M\nonumber\\
&\!\!\!\!&\!\!\!\!+\beta\sum_{\bm{\varphi}_i \in \bm{\Psi}}{\!\!f(\bm{\varphi}_i)V({\mathbf{y}}_{\bm{\varphi}_i}^{**})}\nonumber\\
&\!\!\!\!=&\!\!\!\!\sum_{k=1}^M{y_{m_k'}^*(R_M+C_M)}-MC_M\nonumber\\
&\!\!\!\!&\!\!\!\!+\beta[f(0,0,...,0)V(y_1^{**},...,T^2(p_j),...,y_N^{**})\nonumber\\
&\!\!\!\!&\!\!\!\!+\cdots\nonumber\\
&\!\!\!\!&\!\!\!\!+f(1,1,...,1)V(y_1^{**},...,T^2(p_j),...,y_N^{**})]
\label{eq:eq16}
\end{eqnarray}
where subscript $m'_k$ denotes the index of chosen channel in action $\bm{\alpha}'$, and ${\mathbf{y}}_{\bm{\varphi}_i}^{**}$ denotes the corresponding system belief in the next time slot, and $T^n(p)$ is defined as:
\begin{equation}
T^n(p)=T^{(n-1)}(T(p))=\frac{\lambda_0}{1-\sigma}(1-\sigma^n)+\sigma^np \label{eq:eq17}.
\end{equation}
From (\ref{eq:eq16}) it is clear that $V(\mathbf{y}_{\bm{\varphi}_i}^*)$ will be affine in $p_j$ as soon as the system choose channel $j$ and allocate power to it. If the system keeps not choosing channel $j$ till $n$ goes to infinity, $V(\mathbf{y}_{\bm{\varphi}_i}^*)$ will become $V(c_1,c_2,...,\frac{\lambda_0}{1-\sigma},...,c_N)$ ($c_1,...,c_N$ are constants) since $T^n(p)\to \frac{\lambda_0}{1-\sigma}$ when $n \to \infty$. In this situation, $V(\mathbf{y}_{\bm{\varphi}_i}^*)$ is also affine in $p_j$.\\
\indent From all above, we prove that $V_a(\mathbf{p}), a \in \mathbf{B}$ is affine in $p_j$.
\end{proof}

\indent \textbf{\emph{Lemma 2:}} The value function $V(\mathbf{p})$ is convex in $p_j$ and the following inequality holds:
\begin{eqnarray}
&&V_a(p_1,p_2,...,p_{j-1},cp+(1-c)p',p_{j+1},...,p_N)\nonumber\\
&\leq&cV_a(p_1,p_2,...,p_{j-1},p,p_{j+1},...,p_N) \nonumber\\
&+&(1-c)V_a(p_1,p_2,...,p_{j-1},p',p_{j+1},...,p_N)
\label{eq:eq18}
\end{eqnarray}
\begin{proof}
The inequality holds when $V(\mathbf{p})$ is convex in $p_j$. So we just need to prove the convexity of $V(\mathbf{p})$. Let $V^n(\mathbf{p})$ be the expected reward when the decision horizon spans only $n$ time slots.

When $n=1$, from equation (\ref{eq:eq7}) and (\ref{eq:eq12}), we have:
\begin{eqnarray}
V^1(\mathbf{p})&\!\!\!\!=&\!\!\!\!\max_{\bm{\alpha}\in \mathbf{B}}{\{V^1_{\bm{\alpha}}(\mathbf{p})\}}\nonumber\\
&\!\!\!\!=&\!\!\!\!\max_{\bm{\alpha}\in \mathbf{B}}{\{g_{_{{\bm{\alpha}}}}\!(\mathbf{p})\}}\nonumber\\
&\!\!\!\!=&\!\!\!\!\max_{\bm{\alpha}\in \mathbf{B}}{\bigg\{\sum_{i=1}^{||\bm\alpha||}{p_{m_i}(R_{||\bm\alpha||}+C_{||\bm\alpha||})}-{||\bm\alpha||}C_{||\bm\alpha||}\bigg\}}\nonumber\\
\label{eq:eq19}
\end{eqnarray}

We can easily notice the fact that every element in set $\{\sum_{i=1}^{||\bm\alpha||}{p_{m_i}(R_{||\bm\alpha||}+C_{||\bm\alpha||})}-{||\bm\alpha||}C_{||\bm\alpha||}\}$ is affine and non-decreasing. So $V^1(\mathbf{p})$ is convex in $p_j$.

Next, we assume $V^k(\mathbf{p})$ is convex in $p_j$, $k \geq 1$, and we now prove $V^{k+1}(\mathbf{p})$ is also convex in $p_j$. We have:
\begin{equation}
V^{k+1}(\mathbf{p})=\max_{\bm{\alpha} \in \mathbf{B}}{\{V^{k+1}_{\bm{\alpha}}(\mathbf{p})\}} \label{eq:eq20}
\end{equation}
where
\begin{eqnarray}
V^{k+1}_{\bm{\alpha}}(\mathbf{p})&=&\sum_{i=1}^{||\bm\alpha||}{p_{m_i}(R_{||\bm\alpha||}+C_{||\bm\alpha||})}-{||\bm\alpha||}C_{||\bm\alpha||}\nonumber\\
&&+\beta\sum_{\bm{\varphi}_i \in \bm{\Psi}}{\!\!f(\bm{\varphi}_i)V^k({\mathbf{y}}_{\bm{\varphi}_i}^*)}
\label{eq:eq21}
\end{eqnarray}
The first and second term in equation (\ref{eq:eq21}) are both affine in $p_j$, so they are convex in $p_j$. Next we consider the third term in (\ref{eq:eq21}).
From (\ref{eq:eq11}) and (\ref{eq:eq13}), we know that each element in the third term $\beta\sum_{\bm{\varphi}_i \in \bm{\Psi}}{\!\!f(\bm{\varphi}_i)V^k({\mathbf{y}}_{\bm{\varphi}_i}^*)}$ is either affine in $p_j$ (when $y_j^*=\lambda_0$ or $\lambda_1$) or convex in $p_j$ (when $y_j^*=T(p_j)$). So the third term is also convex in $p_j$. Now we have proved $V^{k+1}_{\bm{\alpha}}(\mathbf{p})$ is also convex in $p_j$.

From all above, we can draw the conclusion that for all $n\geq 1$, $V^n(\mathbf{p})$ is convex in $p_j$. Since $V(\mathbf{p})$ is the infinite form of $V^n(\mathbf{p})$ when $n \to \infty$, so $V(\mathbf{p})$ is convex in $p_j$.
\end{proof}

\indent \textbf{\emph{Lemma 3:}} Suppose a belief vector $\mathbf{p}'=(p'_1,p'_2,...,p'_N)$ is obtained by randomly swapping the positions of the elements in belief vector $\mathbf{p}=(p_1,p_2,...,p_N)\:(0\leq p_j\leq 1)$, the following equality holds: $V(\mathbf{p})=V(\mathbf{p}')$.
\begin{proof}
First, we prove that for all $\bm{\alpha}\in \mathbf{B}$, there exists $\bm{\alpha}'\in \mathbf{B}$ that satisfies $V_{\bm{\alpha}}(\mathbf{p})=V_{\bm{\alpha}'}(\mathbf{p'})$.

For action $\bm{\alpha}$, let $M=||\bm{\alpha}||$ be the number of used channels, $m_1,m_2,...,m_M$ be the channel indexes and $p_{m_1},p_{m_2},...,p_{m_{\!M}}$ be the believes of the used channels. Since $\mathbf{p}$ and $\mathbf{p}'$ have the same elements (in different order), we can find channels $m'_1,m'_2,...,m'_M$ that satisfy the condition that $p_{m_i}=p'_{m'_i}\:(i \in \{1,2,...,M\})$. That is, we can find action $\bm{\alpha}'$ that satisfies $V_{\bm{\alpha}} (\mathbf{p})=V_{\bm{\alpha}'}(\mathbf{p}')$, where $m_i'$ indicates the index of used channel in action $\bm{\alpha}'$.

From above, we can establish a bijection $f:\mathbf{p}\leftrightarrow\mathbf{p'}$ that satisfies $V_{\bm{\alpha}} (\mathbf{p})=V_{\bm{\alpha}'}(\mathbf{p}')$. Consequently, we have $max_{\bm{\alpha}}{\{V_{\bm{\alpha}}(\mathbf{p})\}}=max_{\bm{\alpha}'}{\{V_{\bm{\alpha}'}(\mathbf{p}')\}}$. Therefore, $V(\mathbf{p})=V(\mathbf{p}')$.
\end{proof}
\subsection{Properties of the decision regions of policy $\pi^*$}
\indent Define $\bm{\Phi}_a$ as the decision region of action $a$. That is, action $a$ is optimal when belief is in $\bm{\Phi}_a$.
\begin{equation}
\bm{\Phi}_a=\{\mathbf{p} | V(\mathbf{p})=V_a(\mathbf{p}),a\in \mathbf{B}\}\label{eq:eq22}
\end{equation}

 \textbf{\emph{Definition 1:}} If given $(p_1,...,p_{j-1},x_1,p_{j+1},...,p_N )$, $(p_1,...,p_{j-1},x_2,p_{j+1},...,p_N )\in \bm{\Phi}_a$, $x_1\leq x_2$, $1\leq j\leq N$, $\forall x\in[x_1,x_2]$, we have $(p_1,...,p_{j-1},x,p_{j+1},...,p_N )\in \bm{\Phi}_a$, then we say $\bm{\Phi}_a$ is contiguous along $p_j$ dimension.

\textbf{\emph{Theorem 1:}} $\bm{\Phi}_a$ is contiguous along $p_1,p_2,...,p_N$ dimension $(a \in \mathbf{B})$.
\begin{proof}
Here we prove that $\bm{\Phi}_a$ is contiguous along $p_1$ dimension, the rest can be proved in a similar manner.

Let $(x_1,p_2,...,p_N),(x_2,p_2,...,p_N)\in \bm{\Phi}_a$ and $x_1 \leq x_2$, we \\have $V(x_1,p_2,...,p_N)=V_a(x_1,p_2,...,p_N)$, $V(x_2,p_2,...,p_N)\\=V_a(x_2,p_2,...,p_N)$. $\forall x\in[x_1,x_2]$, $x$ can be expressed as $cx_1+(1-c)x_2$, where $0\leq c\leq 1$.

From lemma 1 and lemma 2, we have:
\begin{eqnarray}
&&V(x,p_2,...,p_N )\nonumber\\
&=&V(cx_1+(1-c)x_2,p_2,...,p_N)\nonumber\\
&\leq&cV(x_1,p_2,...,p_N)+(1-c)V(x_2,p_2,...,p_N)\nonumber\\
&=&cV_a(x_1,p_2,...,p_N)+(1-c)V_a(x_2,p_2,...,p_N)\nonumber\\
&=&V_a(cx_1+(1-c)x_2,p_2,...,p_N)\nonumber\\
&=&V_a(x,p_2,...,p_N)\nonumber\\
&\leq&V(x,p_2,...,p_N )\label{eq:eq23}
\end{eqnarray}

From (\ref{eq:eq23}) we have $V_a(x,p_2,...,p_N)=V(x,p_2,...,p_N)$, that is, $x\in \bm{\Phi}_a$. Therefore $\bm{\Phi}_a$ is contiguous along $p_1$ dimension.
\end{proof}

\subsection{Structure of the optimal policy over 3-dimensional state space}
In order to visually demonstrate the structure of the optimal policy, we consider a system with 3 parallel channels in this section. In this system, each belief is a three-dimensional vector $\mathbf{p} \in \mathbf{O}=([0,1],[0,1],[0,1])$. Each action is also a three-dimensional vector $\bm{\alpha}=(a_1,a_2,a_3), a_j \in \{0,1\}, j \in \{1,2,3\}$. It is clear that there are 8 different actions in total, each has a corresponding decision region. The following theorem summarises the features of each decision region.

\indent \textbf{\emph{Theorem 2:}} $\bm{\Phi}_{(0,0,0)}$ and $\bm{\Phi}_{(1,1,1)}$ are self-symmetric with respect to plane $p_1=p_2$, $p_1=p_3$ and $p_2=p_3$; $\bm{\Phi}_{(0,0,1)}$ and $\bm{\Phi}_{(1,1,0)}$ are self-symmetric with respect to plane $p_1=p_2$; $\bm{\Phi}_{(0,1,0)}$ and $\bm{\Phi}_{(1,0,1)}$ are self-symmetric with respect to plane $p_1=p_3$; $\bm{\Phi}_{(0,1,1)}$ and $\bm{\Phi}_{(1,0,0)}$ are self-symmetric with respect to plane $p_2=p_3$. $\bm{\Phi}_{(1,0,1)}$ and $\bm{\Phi}_{(0,1,1)}$, $\bm{\Phi}_{(1,0,0)}$ and $\bm{\Phi}_{(0,1,0)}$ are mirror-symmetric with respect to plane $p_1=p_2$; $\bm{\Phi}_{(0,0,1)}$ and $\bm{\Phi}_{(1,0,0)}$, $\bm{\Phi}_{(1,1,0)}$ and $\bm{\Phi}_{(0,1,1)}$ are mirror-symmetric with respect to plane $p_1=p_3$; $\bm{\Phi}_{(0,0,1)}$ and $\bm{\Phi}_{(0,1,0)}$, $\bm{\Phi}_{(1,0,1)}$ and $\bm{\Phi}_{(1,1,0)}$ are mirror-symmetric with respect to plane $p_2=p_3$.
\begin{proof}
Let $(p_1,p_2,p_3)\in \bm{\Phi}_{(0,0,0)}$, then we have $V(p_1,p_2,p_3)=V_{(0,0,0)}(p_1,p_2,p_3)$.
From (\ref{eq:eq12}) and lemma 3, we have:
\begin{eqnarray}
&&V_{(0,0,0)}(p_1,p_2,p_3)\nonumber\\
&=&\beta V(T(p_1),T(p_2),T(p_3))\nonumber\\
&=&\beta V(T(p_1),T(p_3),T(p_2))\nonumber\\
&=&\beta V(T(p_2),T(p_1),T(p_3))\nonumber\\
&=&\beta V(T(p_2),T(p_3),T(p_1))\nonumber\\
&=&\beta V(T(p_3),T(p_1),T(p_2))\nonumber\\
&=&\beta V(T(p_3),T(p_2),T(p_1))
\label{eq:eq24}
\end{eqnarray}
That is,
\begin{eqnarray}
&&V_{(0,0,0)}(p_1,p_2,p_3)\nonumber\\
&=&V_{(0,0,0)}(p_1,p_3,p_2)\nonumber\\
&=&V_{(0,0,0)}(p_2,p_1,p_3)\nonumber\\
&=&V_{(0,0,0)}(p_2,p_3,p_1)\nonumber\\
&=&V_{(0,0,0)}(p_3,p_1,p_2)\nonumber\\
&=&V_{(0,0,0)}(p_3,p_2,p_1)
\label{eq:eq25}
\end{eqnarray}
So $\bm{\Phi}_{(0,0,0)}$ is self-symmetric with respect to plane $p_1=p_2$, $p_1=p_3$ and $p_2=p_3$. Similarly we can prove $\bm{\Phi}_{(1,1,1)}$ is self-symmetric with respect to plane $p_1=p_2$, $p_1=p_3$ and $p_2=p_3$.

Next we prove $\bm{\Phi}_{(1,0,0)}$ and $\bm{\Phi}_{(0,1,0)}$ are mirror-symmetric with respect to plane $p_1=p_2$. Let $(p_1,p_2,p_3)\in \bm{\Phi}_{(1,0,0)}$, then $V(p_1,p_2,p_3)=V_{(1,0,0)}(p_1,p_2,p_3)$.
From lemma 3, we have:
\begin{eqnarray}
&\!\!\!\!&\!\!\!\!V(p_2,p_1,p_3)\nonumber\\
&\!\!\!\!=&\!\!\!\!V_{(0,1,0)}(p_2,p_1,p_3)\nonumber\\
&\!\!\!\!=&\!\!\!\!p_1(R_1+C_1)-C_1+\nonumber\\
&\!\!\!\!&\!\!\!\!\beta[p_1V(T(p_2),\lambda_1,T(p_3))+(1-p_1)V(T(p_2),\lambda_0,T(p_3))]\nonumber\\
&\!\!\!\!=&\!\!\!\!p_1(R_1+C_1)-C_1+\nonumber\\
&\!\!\!\!&\!\!\!\!\beta[p_1V(\lambda_1,T(p_2),T(p_3))+(1-p_1)V(\lambda_0,T(p_2),T(p_3))]\nonumber\\
&\!\!\!\!=&\!\!\!\!V_{(1,0,0)}(p_1,p_2,p_3)\nonumber\\
&\!\!\!\!=&\!\!\!\!V(p_1,p_2,p_3)
\label{eq:eq26}
\end{eqnarray}
So $(p_2,p_1,p_3)\in \bm{\Phi}_{(0,1,0)}$, that is, $\bm{\Phi}_{(1,0,0)}$ and $\bm{\Phi}_{(0,1,0)}$ are mirror-symmetric with respect to plane $p_1=p_2$. The rest of the theorem can be proved in a similar way.
\end{proof}


After obtaining the basic features of the decision regions, we now discuss the distribution of the decision regions in the 3-dimension belief space. First we consider the 8 vertices of the cubic belief space:$(0,0,0)$, $(1,0,0)$, $(0,1,0)$, $(0,0,1)$, $(1,1,0)$, $(1,0,1)$, $(0,1,1)$ and $(1,1,1)$. From equation (\ref{eq:eq12}), it is straightforward to obtain the following result:
\begin{equation}
\begin{cases}
 V(0,0,0)=V_{(0,0,0)}(0,0,0)  \\
 V(1,0,0)=V_{(1,0,0)}(1,0,0)  \\
 V(0,1,0)=V_{(0,1,0)}(0,1,0)  \\
 V(0,0,1)=V_{(0,0,1)}(0,0,1)  \\
 V(1,1,0)=V_{(1,1,0)}(1,1,0)  \\
 V(1,0,1)=V_{(1,0,1)}(1,0,1)  \\
 V(0,1,1)=V_{(0,1,1)}(0,1,1)  \\
 V(1,1,1)=V_{(1,1,1)}(1,1,1)
\end{cases}
\Rightarrow \ \
\begin{cases}
 (0,0,0)\in\bm{\Phi}_{(0,0,0)}  \\
 (1,0,0)\in\bm{\Phi}_{(1,0,0)}  \\
 (0,1,0)\in\bm{\Phi}_{(0,1,0)}  \\
 (0,0,1)\in\bm{\Phi}_{(0,0,1)}  \\
 (1,1,0)\in\bm{\Phi}_{(1,1,0)}  \\
 (1,0,1)\in\bm{\Phi}_{(1,0,1)}  \\
 (0,1,1)\in\bm{\Phi}_{(0,1,1)}  \\
 (1,1,1)\in\bm{\Phi}_{(1,1,1)}
\end{cases}
\label{eq:eq27}
\end{equation}

Next, we consider the 12 edges of the belief space cube. We take the plane $p_3=0$ as an example to discuss the four edges on it. When $p_3=0$, we have:
\begin{equation}
\begin{cases}
 V_{(0,0,0)}=&\!\!\!\!\!\beta V(T(p_1),T(p_2),\lambda_0)  \\
 V_{(0,0,1)}=&\!\!\!\!\!-C_1+\beta V(T(p_1),T(p_2),\lambda_0)  \\
 V_{(0,1,0)}=&\!\!\!\!\!p_2(R_1+C_1)-C_1+\beta[p_2V(T(p_1),\lambda_1,\lambda_0)  \\
 &\!\!\!\!\!+(1-p_2)V(T(p_1),\lambda_0,\lambda_0)]  \\
 V_{(1,0,0)}=&\!\!\!\!\!p_1(R_1+C_1)-C_1+\beta[p_1V(\lambda_1,T(p_2),\lambda_0)  \\
 &\!\!\!\!\!+(1-p_1)V(\lambda_0,T(p_2),\lambda_0)]  \\
 V_{(0,1,1)}=&\!\!\!\!\!p_2(R_2+C_2)-2C_2+\beta[p_2V(T(p_1),\lambda_1,\lambda_0)  \\
 &\!\!\!\!\!+(1-p_2)V(T(p_1),\lambda_0,\lambda_0)]  \\
 V_{(1,0,1)}=&\!\!\!\!\!p_1(R_2+C_2)-2C_2+\beta[p_1V(\lambda_1,T(p_2),\lambda_0)  \\
 &\!\!\!\!\!+(1-p_1)V(\lambda_0,T(p_2),\lambda_0)]  \\
 V_{(1,1,0)}=&\!\!\!\!\!(p_1+p_2)(R_2+C_2)-2C_2+\beta[p_1p_2V(\lambda_1,\lambda_1,\lambda_0)  \\
 &\!\!\!\!\!\!+(1-p_1)p_2V(\lambda_0,\lambda_1,\lambda_0)+p_1(1-p_2)V(\lambda_1,\lambda_0,\lambda_0)  \\
 &\!\!\!\!\!\!+(1-p_1)(1-p_2)V(\lambda_0,\lambda_0,\lambda_0)]  \\
 V_{(1,1,1)}=&\!\!\!\!\!(p_1+p_2)(R_3+C_3)-3C_3+\beta[p_1p_2V(\lambda_1,\lambda_1,\lambda_0)  \\
 &\!\!\!\!\!\!+(1-p_1)p_2V(\lambda_0,\lambda_1,\lambda_0)+p_1(1-p_2)V(\lambda_1,\lambda_0,\lambda_0)  \\
 &\!\!\!\!\!\!+(1-p_1)(1-p_2)V(\lambda_0,\lambda_0,\lambda_0)]
\end{cases}
\label{eq:eq28}
\end{equation}
In Section II, we assume that $R_b<R_a<bR_b/a$, $C_b<C_a<bC_b/a$ and $R_a>C_a (1 \leq a \leq b \leq M)$, so we can learn from (\ref{eq:eq28}) that $V_{(0,0,0)}>V_{(0,0,1)}$, $V_{(0,1,0)}>V_{(0,1,1)}$, $V_{(1,0,0)}>V_{(1,0,1)}$, $V_{(1,1,0)}>V_{(1,1,1)}$. Therefore, the optimal actions on this plane are restricted to the following four actions: $(0,0,0),(0,1,0),(1,0,0),(1,1,0)$.

\begin{figure}[!t]
\centering
\subfloat[$p_1=0$]{\includegraphics[width=1.6in]{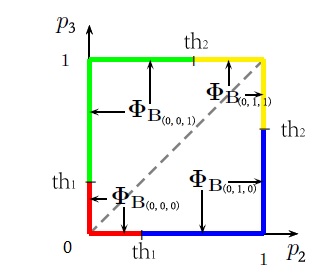}
\label{fig1.a}}
\hfil
\subfloat[$p_1=1$]{\includegraphics[width=1.6in]{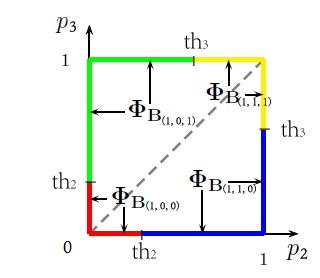}
\label{fig1.b}}
\hfil
\subfloat[$p_2=0$]{\includegraphics[width=1.6in]{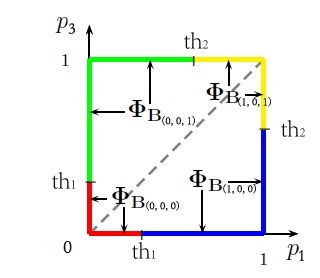}
\label{fig1.c}}
\hfil
\subfloat[$p_2=1$]{\includegraphics[width=1.6in]{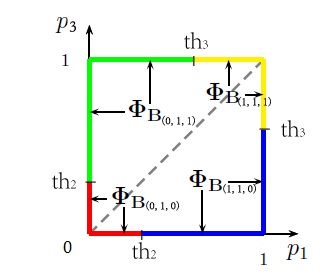}
\label{fig1.d}}
\hfil
\subfloat[$p_3=0$]{\includegraphics[width=1.6in]{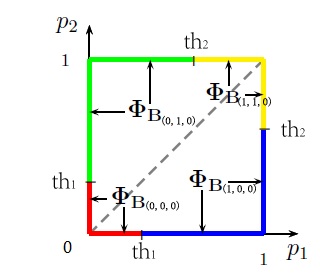}
\label{fig1.e}}
\hfil
\subfloat[$p_3=1$]{\includegraphics[width=1.6in]{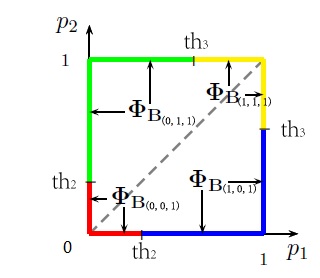}
\label{fig1.f}}
\caption{Structure of the optimal policy on the boundary}
\label{fig1}
\end{figure}

On edge $\{p_1=0,p_3=0\}$, according to lemma 2 and the assumption in Section II, we have: $V_{(0,1,0)}>V_{(1,1,0)}$ and $V_{(0,1,0)}>V_{(1,0,0)}$. With this we know the optimal action on this edge is either $(0,1,0)$ or $(0,0,0)$. From (\ref{eq:eq28}) we have:
\begin{eqnarray}
V_{(0,1,0)}\!\!-V_{(0,0,0)}&\!\!\!\!\!=&\!\!\!\!\!p_2(R_2+C_2)-2C_2+\beta[p_2V(\lambda_0,\lambda_1,\lambda_0)\nonumber\\
&\!\!\!\!\!+&\!\!\!\!\!(1-p_2)V(\lambda_0,\lambda_0,\lambda_0)-V(\lambda_0,T(p_2),\lambda_0)]\nonumber\\
\label{eq:eq29}
\end{eqnarray}
Due to the convexity of $V(\mathbf{p})$, there exists

\begin{equation} th_1=\frac{C_1+\beta[V(\lambda_0,T(p_2),\lambda_0)-V(\lambda_0,\lambda_0,\lambda_0)]}{R_1+C_1+V(\lambda_0,\lambda_1,\lambda_0)-V(\lambda_0,\lambda_0,\lambda_0)}
\label{eq:eq30}
\end{equation}
so that when $p_2\geq Th_1$, $(0,p_2,0)\in \bm{\Phi}_{(0,1,0)}$; when $p_2\leq Th_1$, $(0,p_2,0)\in \bm{\Phi}_{(0,0,0)}$ (Fig. \ref{fig1.e}).

For the edge $p_1=1,p_3=0$ (Fig. \ref{fig1.e}), in the same manner we have $V_{(1,1,0)}>V_{(0,1,0)}$ and $V_{(1,1,0)}>V_{(0,0,0)}$. Thus, the optimal action on this edge is either $(1,1,0)$ or $(1,0,0)$. From (\ref{eq:eq28}) we have:
\begin{eqnarray}
V_{B_{(1,1,0)}}\!\!-V_{B_{(1,0,0)}}&\!\!\!\!\!=&\!\!\!\!\!(1+p_2)(R_2+C_2)-2C_2-R_1\nonumber\\
&\!\!\!\!\!+&\!\!\!\!\!\beta[p_2V(\lambda_1,\lambda_1,\lambda_0)+(1-p_2)V(\lambda_1,\lambda_0,\lambda_0)\nonumber\\
&\!\!\!\!\!-&\!\!\!\!\!V(\lambda_1,T(p_2),\lambda_0)]
\label{eq:eq31}
\end{eqnarray}
Due to the convexity of $V(\mathbf{p})$, there exists
\begin{equation} th_2=\frac{R_1-R_2+C_2+\beta[V(\lambda_1,T(p_2),\lambda_0)-V(\lambda_1,\lambda_0,\lambda_0)]}{R_2+C_2+V(\lambda_1,\lambda_1,\lambda_0)-V(\lambda_1,\lambda_0,\lambda_0)}
\label{eq:eq32}
\end{equation}
so that when $p_2\geq Th_2$, $(1,p_2,0)\in \bm{\Phi}_{(1,1,0)}$, when $p_2\leq Th_2$, $(1,p_2,0)\in \bm{\Phi}_{(0,1,0)}$.

Using the symmetric properties in Theorem 2, we can easily derive similar results on the other planes and edges. So the structure of the optimal policy on the 6 planes of the cubic belief space is shown in Fig. \ref{fig1}, where
\begin{equation}
\begin{cases}
th_1=\frac{C_1+\beta[V(T(th_1),\lambda_0,\lambda_0)-V(\lambda_0,\lambda_0,\lambda_0)]}{R_1+C_1+V(\lambda_0,\lambda_1,\lambda_0)-V(\lambda_0,\lambda_0,\lambda_0)}\\
th_2=\frac{R_1-R_2+C_2+\beta[V(T(th_2),\lambda_1,\lambda_0)-V(\lambda_1,\lambda_0,\lambda_0)]}{R_2+C_2+V(\lambda_1,\lambda_1,\lambda_0)-V(\lambda_1,\lambda_0,\lambda_0)}\\
th_3=\frac{2R_2-2R_3+C_3+\beta[V(T(th_3),\lambda_1,\lambda_1)-V(\lambda_1,\lambda_1,\lambda_0)]}{R_3+C_3+V(\lambda_1,\lambda_1,\lambda_1)-V(\lambda_1,\lambda_1,\lambda_0)}
\end{cases}
\label{eq:eq33}
\end{equation}

After the threshold on each edge is found, we next derive the structure of the optimal policy in the whole cube.

\begin{figure}[!t]
\centering
\subfloat[Belief space region segmentation]{\includegraphics[width=2.6in]{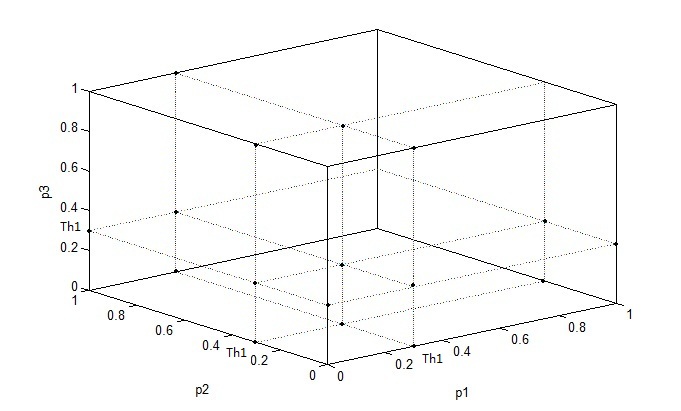}
\label{fig2.a}}
\hfil
\subfloat[$V_a(x,0,z)$]{\includegraphics[width=1.6in]{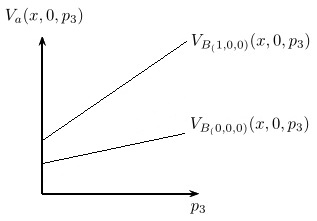}
\label{fig2.b}}
\hfil
\subfloat[$V_a(x,y,z)$]{\includegraphics[width=1.6in]{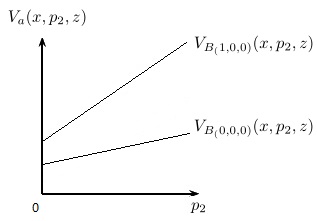}
\label{fig2.c}}
\hfil
\caption{Belief Space Segmentation}
\label{fig2}
\end{figure}

\textbf{\emph{Theorem 3:}} $\bm{\Phi}_a$ is a simple connected region extended from the vertices $v_a$ of the cubic belief space $([0,1],[0,1],[0,1])$, where
\begin{equation}
v_a=\begin{cases}
(0,0,0)&a=(0,0,0) \\
(1,0,0)&a=(1,0,0) \\
(0,1,0)&a=(0,1,0) \\
(0,0,1)&a=(0,0,1) \\
(1,1,0)&a=(1,1,0) \\
(1,0,1)&a=(1,0,1) \\
(0,1,1)&a=(0,1,1) \\
(1,1,1)&a=(1,1,1)
\end{cases}
\label{eq:eq34}
\end{equation}
\begin{proof}
From (\ref{eq:eq27}) we already have $v_a\in\bm{\Phi}_a$, and from Theorem 1 we know $\bm{\Phi}_a$ has \emph{at least} one connected region extended from $v_a$. Thus here we only need to prove that $\bm{\Phi}_a$ has \emph{only} one connected region.

Take $\bm{\Phi}_{(0,0,0)}$ as an example. Let $\bm{\Phi}'_a$ be a connected region extended from $(0,0,0)$. Because of the symmetry of the region, there is a minimum cube $([0,Th1],[0,Th1],[0,Th1])$ that includes $\bm{\Phi}'_a$, as shown in Fig. \ref{fig2.a}, and the state space are split into several cubes. Due to the minimality of cube $([0,Th1],[0,Th1],[0,Th1])$, we have $Th_1>th_1$.

Consider the cube $([0,Th_1],[0,Th_1],[0,1])$, suppose there exists another region $\bm{\Phi}''_a$ in it, then $\forall (x,y,z)\in\bm{\Phi}''_a$, line $p_1=p_2=Th_1$ will pass across both $\bm{\Phi}'_a$ and $\bm{\Phi}''_a$, which makes $\bm{\Phi}'_a$ and $\bm{\Phi}''_a$ connected. Therefore, no such region $\bm{\Phi}''_a$ exists in cube $([0,Th_1],[0,Th_1],[0,1])$. Similarly, we can prove there exists no $\bm{\Phi}''_a$ in cube $([0,Th_1],[0,1],[0,Th_1])$ or $([0,1],[0,Th_1],[0,Th_1])$.

Next we consider the cube $([Th_1,1],[0,1],[0,1])$. $\forall v=(x,y,z)\in([Th_1,1],[0,1],[0,1])$, since $Th_1 >th_1$, we have $V_{(1,0,0)}(x,0,0)>V_{(0,0,0)}(x,0,0)$. From equation (\ref{eq:eq12}) and Lemma 1, we have:

\begin{equation}
\begin{cases}
\!\!\!\!\!\!&\frac{\partial V_{B_{0,0,0}}(x,0,p_3)}{\partial p_3}=\beta\frac{\partial V(T(x),\lambda_0,T(p_3))}{\partial p_3}\\
\!\!\!\!\!\!&\frac{\partial V_{B_{1,0,0}}(x,0,p_3)}{\partial p_3}=\beta\frac{\partial[xV(\lambda_1,\lambda_0,T(p_3))+(1-x)V(\lambda_0,\lambda_0,T(p_3))]}{\partial p_3}
\end{cases}
\label{eq:eq35}
\end{equation}

From (\ref{eq:eq35}) we have $\frac{\partial V_{B_{0,0,0}}(x,0,p_3)}{\partial p_3}<\frac{\partial V_{B_{1,0,0}}(x,0,p_3)}{\partial p_3}$, so we can tell from Fig. \ref{fig2}(b) that $\forall \ 0 \leq z \leq 1$, $V_{(1,0,0)}(x,0,z)>V_{(0,0,0)}(x,0,z)$. Likewise, we have:

\begin{equation}
\begin{cases}
\!\!\!\!\!\!&\frac{\partial V_{B_{0,0,0}}(x,p_2,z)}{\partial p_2}=\beta\frac{\partial V(T(x),T(p_2),T(z))}{\partial p_2}\\
\!\!\!\!\!\!&\frac{\partial V_{B_{1,0,0}}(x,p_2,z)}{\partial p_2}=\beta\frac{\partial[xV(\lambda_1,T(p_2),T(z))+(1-x)V(\lambda_0,T(p_2),T(z))]}{\partial p_2}
\end{cases}
\label{eq:eq36}
\end{equation}
From (\ref{eq:eq36}) we have $\frac{\partial V_{B_{1,0,0}}(x,p_2,z)}{\partial p_2}>\frac{\partial V_{B_{0,0,0}}(x,p_2,z)}{\partial p_2}$, and from Fig. \ref{fig2}(c) we can tell that $\forall \ 0\!\!\leq\!\!y\!\!\leq\!\!1$, $V_{(1,0,0)}(x,y,z)>V_{(0,0,0)}(x,y,z)$. Therefore, $\forall \ v=(x,y,z)\in([Th_1,1],[0,1],[0,1])$, we have $v\notin\bm{\Phi}_{(0,0,0)}$, that is, there exists no connected region $\bm{\Phi}''_{B_{(0,0,0)}}$ in cube $([Th_1,1],[0,1],[0,1])$. In the same manner, we can also prove that there exists no connected region $\bm{\Phi}''_{(0,0,0)}$ in cube $([0,1],[Th_1,1],[0,1])$ and $([0,1],[0,1],[Th_1,1])$. Now we have proved that there exists on other connected region $\bm{\Phi}''_{(0,0,0)}$ in the whole belief space cube.

The other 7 regions $\bm{\Phi}_{1,0,0}$, $\bm{\Phi}_{(0,1,0)}$, $\bm{\Phi}_{(0,0,1)}$, $\bm{\Phi}_{(1,1,0)}$, $\bm{\Phi}_{(1,0,1)}$, $\bm{\Phi}_{(0,1,1)}$ and $\bm{\Phi}_{(1,1,1)}$ can be proved in the same way.
\end{proof}

\section{Simulation Based on Linear Programming}
\indent Linear programming is one of the approaches to solve the Bellman equation. Based on \cite{cite11}, we model our problem as the following linear programming formulation:
\begin{eqnarray}
&\!\!\!\!\!&\!\!\!\!\!\forall \mathbf{p}\in\mathbb{X}, \forall a\in\mathbb{A}_{\mathbf{p}},\nonumber\\
&\!\!\!\!\!&\!\!\!\!\!\min{\sum_{\mathbf{p}\in\mathbb{X}}V(\mathbf{p})},\quad s.t. \; g_a(\mathbf{p})+\beta\sum_{\mathbf{y}\in\mathbb{X}}{f_a(\mathbf{p},\mathbf{y})V(\mathbf{y})}\leq V(\mathbf{p})\nonumber\\
\label{eq:eq37}
\end{eqnarray}
where $\mathbb{X}$ denotes the belief space, $\mathbb{A}_{\mathbf{p}}$ is the set of available actions for belief state $\mathbf{p}$. The state transition probability $f_a(\mathbf{p},\mathbf{y})$ is the probability that the next state will be $\mathbf{y}$ when the current state is $\mathbf{p}$ and the current action is $a\in\mathbb{A}_{\mathbf{p}}$. The optimal policy is given by
\begin{equation}
\pi(\mathbf{p})=\arg\,\max_{a\in\mathbb{A}_{\mathbf{p}}}{\big(g_a(\mathbf{p})+\beta\sum_{\mathbf{y}\in\mathbb{X}}{f_a(\mathbf{p},\mathbf{y})V(\mathbf{y})}\big)}
\label{eq:eq38}
\end{equation}

\indent For ease of discussion and demonstration, we consider the case of three-dimensional belief space. We use the LOQO solver on NEOS Server \cite{cite12} with AMPL input \cite{cite13} to obtain the solution of equation (\ref{eq:eq37}). Then we use MATLAB to construct the policy according to equation (\ref{eq:eq38}).
\begin{figure}[!t]
\centering
\includegraphics[width=3.6in]{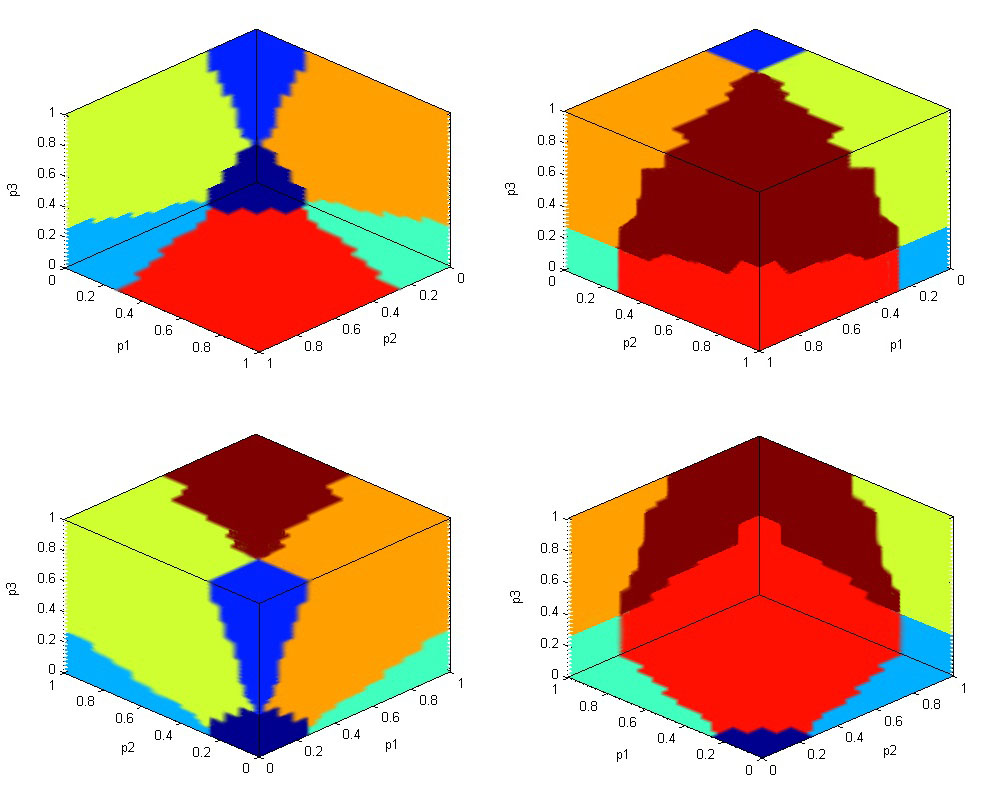}
\caption{Structure of optimal policy.}
\label{fig3}
\end{figure}

\begin{figure}[!t]
\centering
\subfloat[$\bm{\Phi}_{(0,0,0)}$]{\includegraphics[width=1.6in]{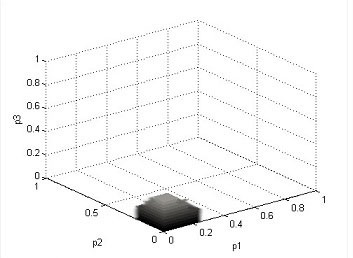}
\label{fig4.a}}
\hfil
\subfloat[$\bm{\Phi}_{(1,0,0)}$]{\includegraphics[width=1.6in]{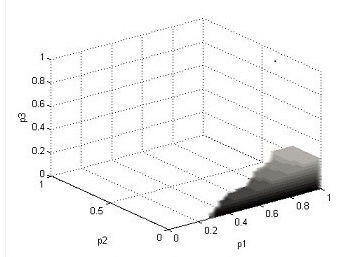}
\label{fig4.b}}
\hfil
\subfloat[$\bm{\Phi}_{(0,1,0)}$]{\includegraphics[width=1.6in]{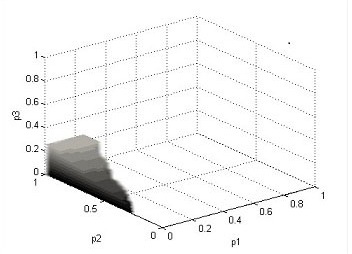}
\label{fig4.c}}
\hfil
\subfloat[$\bm{\Phi}_{(0,0,1)}$]{\includegraphics[width=1.6in]{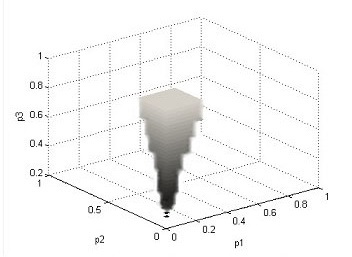}
\label{fig4.d}}
\hfil
\subfloat[$\bm{\Phi}_{(1,1,0)}$]{\includegraphics[width=1.6in]{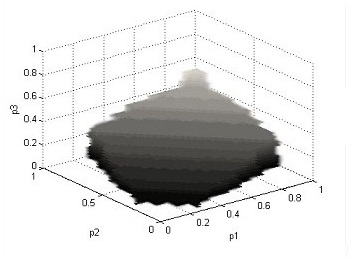}
\label{fig4.e}}
\hfil
\subfloat[$\bm{\Phi}_{(1,0,1)}$]{\includegraphics[width=1.6in]{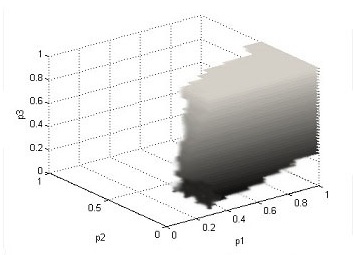}
\label{fig4.f}}
\hfil
\subfloat[$\bm{\Phi}_{(0,1,1)}$]{\includegraphics[width=1.6in]{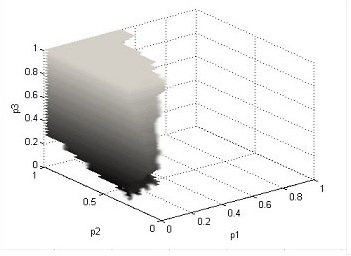}
\label{fig4.g}}
\hfil
\subfloat[$\bm{\Phi}_{(1,1,1)}$]{\includegraphics[width=1.6in]{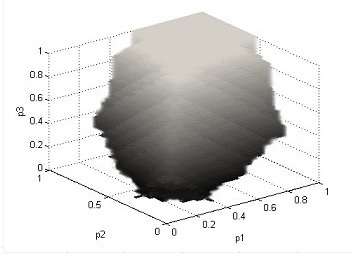}
\label{fig4.h}}
\caption{Individual Decision Regions }
\label{fig4}
\end{figure}

\indent Fig. \ref{fig3} shows the AMPL solution of the value function and the corresponding optimal policy. We use
the following set of parameters: $\lambda_1=0.9$, $\lambda_0=0.1$, $\beta=0.9$, $R_1=3$, $R_2=2$, $R_3=1.78$, $C_1=1.5$, $C_2=1$, $C_3=0.89$. Fig. {\ref{fig4}} shows each of the 8 individual decision regions. We can see clearly in the figure that the decision regions have the symmetry and contiguity properties we gave in Section III.

\indent To better understand the optimal policy, we next investigate how the parameters $\lambda_0, \lambda_1, R_1, R_2, R_3, C_1, C_2, C_3$ affect the structure of the decision regions.

\indent First, we consider the effect of $\lambda_0$ and $\lambda_1$. Let $|\bm{\Phi}_a|$ denote the volume of $\bm{\Phi}_a$ , define the normalized volume $|\bm{\Phi}_a|/|\mathbb{X}|$ as  the volume of $\bm{\Phi}_a$ normalized against the volume of the total belief space $\mathbb{X}$ . Due to the symmetry property of the decision regions, we only study the decision regions for the following 4 actions $(0,0,0),(1,0,0), (1,1,0), (1,1,1)$. For ease of notation, in the following discussion we use $B_0$, $B_1$, $B_2$ and $B_3$ to denote these four actions, respectively.

\begin{figure}[!t]
\centering
\subfloat[$\lambda_0$]{\includegraphics[width=2.8in]{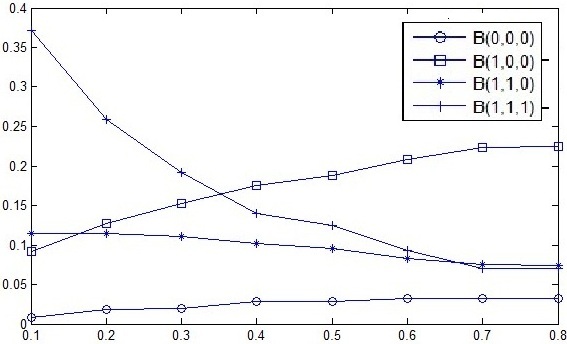}
\label{fig5.a}}
\hfil
\subfloat[$\lambda_1$]{\includegraphics[width=2.8in]{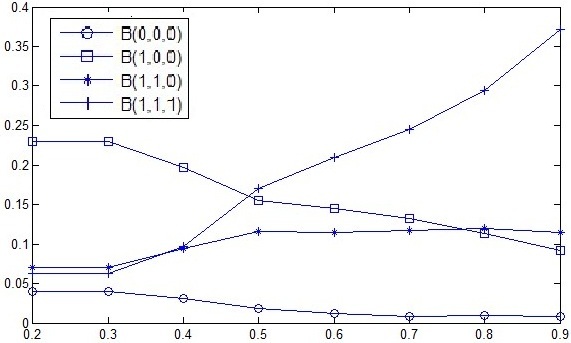}
\label{fig5.b}}
\caption{Normalized $|\bm{\Phi}_a|$ with varying $\lambda_0$, $\lambda_1$ ($R_1=3, R_2=1.75, R_3=1.361, \frac{R_k}{C_k}=2$)}
\label{fig5}
\end{figure}

 We first fix the value of $\lambda_1$ and increase $\lambda_0$ from 0.1 to 0.8. Fig. \ref{fig5.a} shows the normalized volume of the four decision regions with increasing $\lambda_0$. We can see that initially when $\lambda_0=0.1$, $\bm{\Phi}_{B_3}$ has the biggest volume, it then decreases rapidly when $\lambda_0$ increases. The volume of $\bm{\Phi}_{B_1}$ also changes significantly with increasing $\lambda_0$, but in contrast to $\bm{\Phi}_{B_3}$, it increases rapidly when $\lambda_0$ increases. When $\lambda_0=0.8$, $\bm{\Phi}_{B_1}$ has the biggest volume. This trends have the following implications: when $\lambda_0$ is small, which means the channels tend to remain in the bad state, it is beneficial to allocate power to all the channels (choose action $B_3=(1,1,1)$), whilst when $\lambda_0$ is large, which means the channel is very likely to change from bad state to good state, it is better to ``gamble'' on one channel (choose action $B_1=(1,0,0)$).

 Similar trends can be observed in Fig. \ref{fig5.b} which shows the volumes of the four decision regions versus $\lambda_1$. When $\lambda_1$ is small, $\bm{\Phi}_{B_1}$ has the biggest value, which means it is optimal to ``bet'' on one channel when $\lambda_1$ is small. When $\lambda_1$ is greater than $0.49$, $\bm{\Phi}_{B_3}$ overtakes $\bm{\Phi}_{B_1}$, which means when $\lambda_1$ is big enough it is better for the system to take a more conservative action by allocating power to all the channels instead of ``gambling'' on one channel. The interesting thing is that $|\bm{\Phi}_{B_0}|$ and $|\bm{\Phi}_{B_2}|$ change only slightly with varying $\lambda_0$ and $\lambda_1$. This implies that in order to maximize the reward, the system should either allocate the transmission power to all the channels or gamble on one channel. Using part of the channels ($B_2=(1,1,0)$) or doing nothing ($B_0=(0,0,0)$) is always not a good idea to maximize the long term reward. 

\begin{figure}[ht]
\centering
\subfloat[$B_0$]{\includegraphics[width=1.6in]{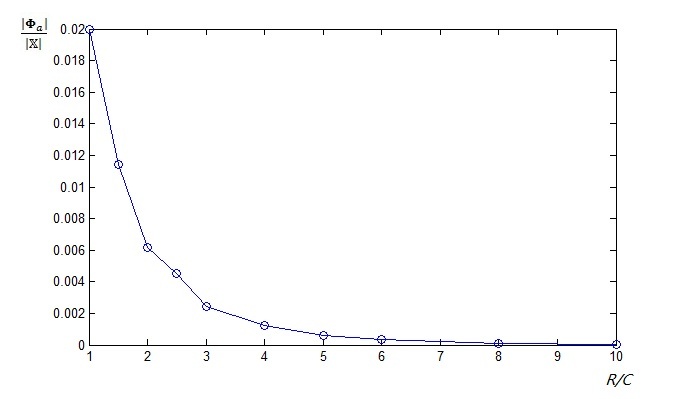}
\label{fig6.a}}
\hfil
\subfloat[$B_1$]{\includegraphics[width=1.6in]{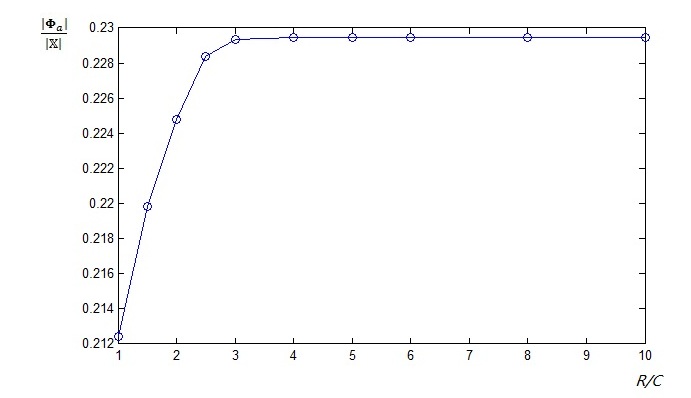}
\label{fig6.b}}
\hfil
\subfloat[$B_2$]{\includegraphics[width=1.6in]{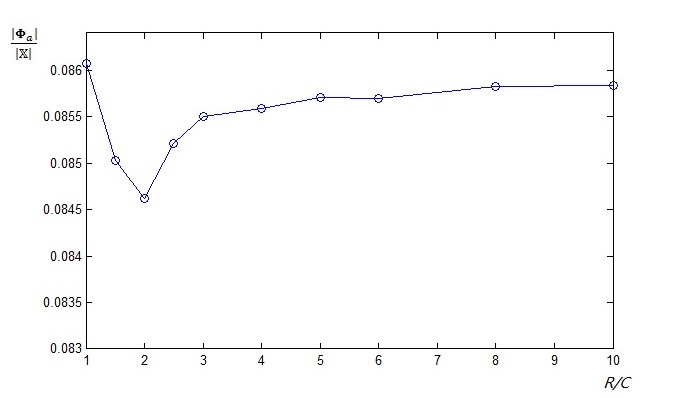}
\label{fig6.c}}
\hfil
\subfloat[$B_3$]{\includegraphics[width=1.6in]{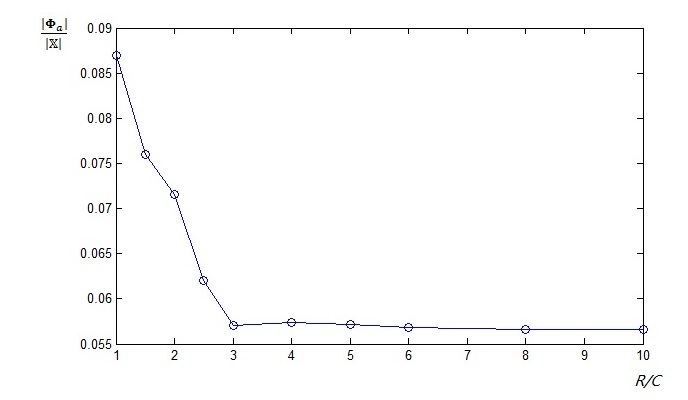}
\label{fig6.d}}
\hfil
\caption{Normalized $|\bm{\Phi}_a|$ vs. $R_k/C_k\:(1\leq k\leq N)$ ($R_1=3,R_2=1.55,R_3=1.06,\lambda_0=0.1,\lambda_1=0.9$)}
\label{fig6}
\end{figure}

\begin{figure}[ht]
\centering
\subfloat[$B_0$]{\includegraphics[width=1.6in]{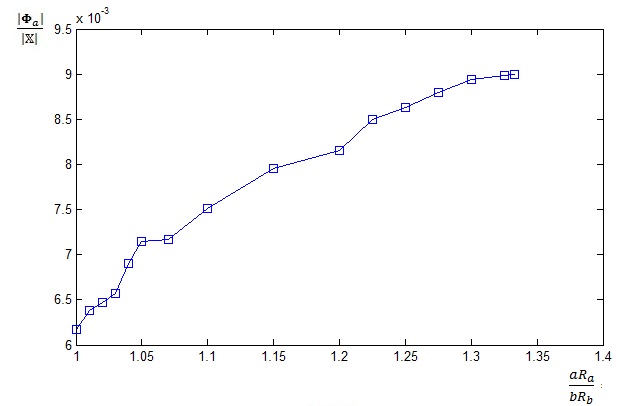}
\label{fig7.a}}
\hfil
\subfloat[$B_1$]{\includegraphics[width=1.6in]{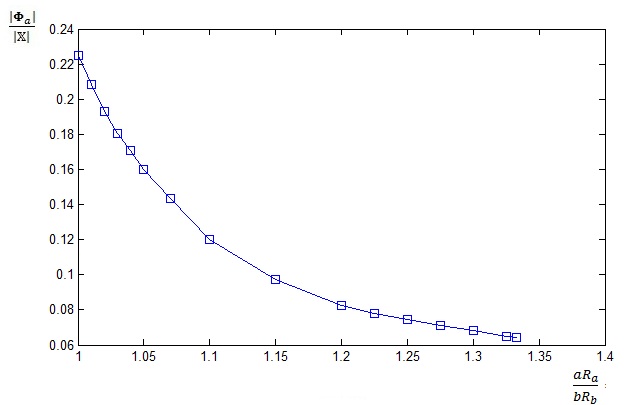}
\label{fig7.b}}
\hfil
\subfloat[$B_2$]{\includegraphics[width=1.6in]{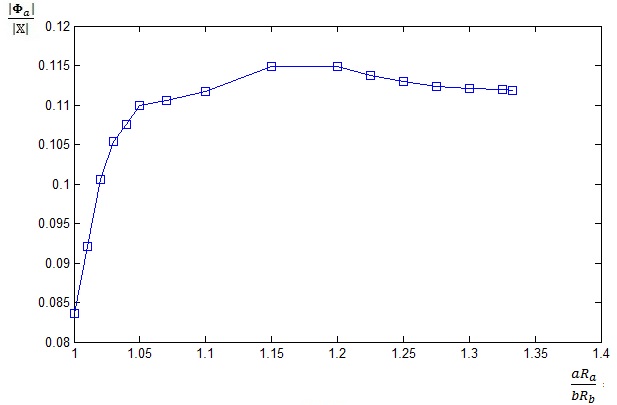}
\label{fig7.c}}
\hfil
\subfloat[$B_3$]{\includegraphics[width=1.6in]{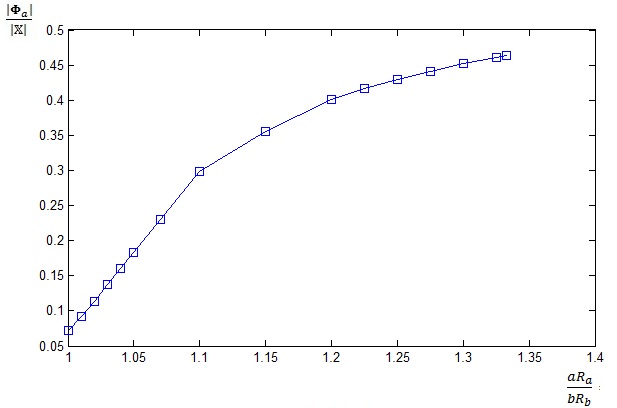}
\label{fig7.d}}
\hfil
\caption{Normalized $|\bm{\Phi}_a|$ with changing $\frac{k_2R_{k_2}}{k_1R_{k_1}}$ ($\lambda_0=0.1,\lambda_1=0.9,R_1=3,C_1=1.5$)}
\label{fig7}
\end{figure}

\indent Next we study the effect of immediate reward $R_k$ and immediate loss $C_k\:(1\leq k\leq N)$ on the structure of the optimal policy. It is straightforward to think that if the ratio of $R_k/C_k$ is large, the total system reward will be large. Fig. \ref{fig6} shows that when $R_k/C_k$ grows, the normalized volume of $\bm{\Phi}_{B_0}$ and $\bm{\Phi}_{B_3}$ decreases, whist $\bm{\Phi}_{B_1}$ grows with $R_k/C_k$. $\bm{\Phi}_{B_2}$ decreases at first and then increases. For all four actions, the volumes of the decision regions reach a constant level respectively and remain unchanged when $R_k/C_k$ grows beyond a certain value. 
\\
\indent In fact, we notice in Fig. \ref{fig6} that the value of $R_k/C_k$ have limited effect on the decision regions in terms of percentage of each decision region in the whole belief space. Now we consider the value of $\frac{k_2R_{k_2}}{k_1R_{k_1}}$ and $\frac{k_2C_{k_2}}{k_1C_{k_1}}$ , and try to find out how they affect the structure of optimal policy (here we fix the value of $R_k/C_k$, so $\frac{k_2C_{k_2}}{k_1C_{k_1}}$ changes along with $\frac{k_2R_{k_2}}{k_1R_{k_1}}$ in the same manner). As in Section III, we assume that $R_{k_2}<R_{k_1}<k_2R_{k_2}/k_1$, $C_{k_2}<C_{k_1}<k_2C_{k_2}/k_1$ and $R_k>C_k (1 \leq k_1 \leq k_2 \leq M)$, so that when more channels are chosen in an action, the system obtains larger immediate reward $kR_k$, therefore our power allocation scheme encourages the system to allocate power to more channels. It is shown in Fig. \ref{fig7} that when $\frac{k_2R_{k_2}}{k_1R_{k_1}}$ grows, normalized $|\bm{\Phi}_{B_1}|$ decreases whilst $|\bm{\Phi}_{B_3}|$ increases. Therefore when $\frac{k_2R_{k_2}}{k_1R_{k_1}}$ is large, the total immediate reward is large enough for the system to act conservatively by allocating the transmission power to all the channels. Whilst when $\frac{k_2R_{k_2}}{k_1R_{k_1}}$ is small, the total immediate reward is so small that system would rather ``gamble'' on one channel. Like the observation in Fig.\ref{fig6}, the values of $|\bm{\Phi}_{B_0}|$ and $|\bm{\Phi}_{B_2}|$ only change slightly with varying $\frac{k_2R_{k_2}}{k_1R_{k_1}}$.

\indent From the discussion above, we can draw a conclusion that: when $\lambda_1-\lambda_0$ and $\frac{k_2R_{k_2}}{k_1R_{k_1}}$ are large, the system tends to act conservatively and share power among all the channels; when $\lambda_1-\lambda_0$ and $\frac{k_2R_{k_2}}{k_1R_{k_1}}$ are small, the system tends to ``gamble'' on one channel. No matter how the parameters change, action $B_2$ is a mediocre choice and bring medium reward thus this action is not often taken. Action $B_0$ is seldom chosen by the system since it brings no immediate reward, it is chosen only when the belief is so small that the system is almost sure to suffer loss.

\section{Conclusion}
In this paper, we have studied the power allocation problem over $N (N \geq 3)$ Gilbert-Elliott channels. We have theoretically derived the threshold-based structure of the optimal policy for $N=3$, and graphically illustrated the structure by formulating and solving a linear programming formulation of the problem. For $N>3$, it is difficult to demonstrate the results graphically, but it is possible to derive the structure mathematically, and we will work on this issue in the future. For future work, we would also like to investigate the case of non-identical channels and use a multi-armed bandit (MAB) formulation to find the thresholds for multiple channel system with $N>3$.


\section*{Acknowledgment}

This work is partially supported by National Key Basic Research Program of China under grant 2013CB329603, Natural Science Foundation of China under grant 61071081 and 60932003. This research was also sponsored in part by the U.S. Army Research Laboratory under the Network Science Collaborative Technology Alliance, Agreement Number W911NF-09-2-0053, and by the
Okawa Foundation, under an Award to support research on ¡°Network Protocols that Learn¡±.



%

\end{document}